\documentclass[12pt,english]{revtex4-1}
\usepackage[T1]{fontenc}
\usepackage[latin9]{inputenc}
\usepackage{geometry}
\usepackage{xcolor}
\usepackage{pdfcolmk}
\usepackage{float}
\usepackage{enumitem}
\usepackage{amsmath}
\usepackage{amssymb}
\usepackage{cancel}
\usepackage{graphicx}
\PassOptionsToPackage{normalem}{ulem}
\usepackage{ulem}

\makeatletter

\newcommand{\lyxdot}{.}

\providecolor{lyxadded}{rgb}{0,0,1}
\providecolor{lyxdeleted}{rgb}{1,0,0}

\DeclareRobustCommand{\lyxdeleted}[3]{{\color{lyxdeleted}\sout{#3}}}

\makeatother

\usepackage{babel}
\begin{document}

\title{Driving Spin-Boson Models From Equilibrium Using Exact Quantum Dynamics}

\author{G.M.G. McCaul, C.D. Lorenz and L. Kantorovich}

\address{Physics Department, King's College London, The Strand, London, WC2R
2LS, United Kingdom}
\begin{abstract}
We present an application of the Extended Stochastic Liouville Equation
(ESLE) {[}\emph{Phys. Rev. B} \textbf{95}, 125124{]}, which gives
an exact solution for the reduced density matrix of an open system
surrounded by a harmonic heat bath. This method considers the \emph{extended}
system (the open system and the bath) being thermally equilibrated
prior to the action of a time dependent perturbation, as opposed to
the usual assumption that system and bath are initially \emph{partitioned}.
This is an exact technique capable of accounting for arbitrary parameter
regimes of the model. Here we present our first numerical implementation
of the method in the simplest case of a Caldeira-Leggett representation
of the bath Hamiltonian, and apply it to a spin-boson system driven
from coupled equilibrium. We observe significant behaviours in both
the transient dynamics and asymptotic states of the reduced density
matrix not present in the usual approximation. 
\end{abstract}
\maketitle

\section{Introduction}

The coupling of an environment to a system introduces phenomena not
found in isolation. In particular, dissipation, Brownian fluctuations
and (in the quantum case) decoherence cannot be observed or explicated
without recourse to environments \citep{Leggett1987}. It is an essential
component of quantum thermodynamics \citep{PhysRevB.91.224303,PhysRevA.95.022120},
and vital to the field of quantum computing, where coherence is a
resource which the environment can dissipate \citep{PhysRevLett.99.267202,PhysRevB.84.245109,Schoelkopf2008}\emph{
}or \emph{enhance }(with suitable environmental engineering) \citep{PhysRevA.82.052112}.

Historically, methods based on the Feynman-Vernon influence functional
\citep{Feynman1963} have had great success in the treatment of environmental
effects \citep{Caldeira1983,Grabert1988,Hanggi1987,Tsusaka1999a,Makri2014,Makri1998,Dattani2012,Carrega2015}.
In this formalism, the extended system-environment Hamiltonian (consisting
of the contributions due to the open system, the environment and their
coupling) is cast as a path integral, where the environment can be
analytically integrated over to give an expression for the reduced
density matrix, representing the open system purely in terms of the
open system operators. This results in a modified propagator for the
open system, where the effect of the environment and its interaction
with the open system is captured by additional terms in the path integral
exponent known as the influence functional. In this approach the inital
density matrix is factorised, i.e. it is assumed that the open system
and its environment are initially \emph{partitioned} and hence there
is no interaction between them at the initial time \citep{Weiss-1999}.

From this representation, a number of techniques have been developed.
A few examples are the Stochastic Schrodinger Equation \citep{Orth2013,Orth2010},
Stochastic Liouville-von Neumann Equation \citep{Stockburger2004}
and the quasiadiabatic path integral \citep{Nalbach2009}. While these
techniques have been applied to a broad range of problems, the spin-Boson
model has proved a particularly popular test-bed \citep{PhysRevLett.116.240403,PhysRevB.82.144423,PhysRevB.91.224303,Orth2013,Orth2010,DemichevChaiChian}.
This model has also been interrogated by other methods, including
cumulant expansions \citep{Reichman1997}, matrix products \citep{Wall2016}
and reaction-coordinate \citep{PhysRevE.95.032139} approaches. It
is also amenable to analytic derivations, both perturbatively \citep{Kayanuma1998}
and nonpertubatively \citep{Dylewsky2016}, as well as to an application
of the Born approximation \citep{DiVincenzo2005}. The model may also
draw on the extensive work done on driven two-state models, most famously
the Landau-Zener sweep \citep{Zener696,Wittig2005,Rojo2010} and its
generalisations \citep{Berman2005,Bambini1981,Carroll1990}. 

In addition, the spin-Boson model displays rich, non-trivial behaviour,
with integrable and non-integrable regimes \citep{Stepanov2008},
diffusive and localised phases \citep{Guinea1985}, as well as coherent
to incoherent crossovers \citep{PhysRevLett.80.4370,PhysRevE.55.R3809}.
Besides the model's obvious application to qubit behaviour, it has
been mapped to impurities in an electronic bath (i.e. Kondo model)
\citep{Vojta2006,Blume1970}, Josephson junctions \citep{Liu2002,RevModPhys.73.357},
cold atoms \citep{PhysRevA.77.051601,PhysRevB.82.144423}, and even
biological systems \citep{doi:10.1063/1.449017}. The spin-Boson model
therefore serves as an excellent toy model, with application to real
experimental systems. 

The main assumption made in the work mentioned above is that initially
the extended system is partitioned, i.e. the open system and bath
are initially non-interacting. The interaction betwen the open system
and bath is then turned on for the dynamical evolution. This, so-called
\emph{partitioned} approach, strictly speaking, is only applicable
for weak system-bath coupling, as well as when studying the long-time
behaviour when transient effects due to the unphysical initial preparation
of the system die away. Our recent work introduced the Extended Stochastic
Liouville Equation (ESLE) method \citep{ourpaper}, which enables
one to project out the environment \emph{exactly}, without assuming
that it is decoupled from the open system at the initial time. The
ESLE is a non-perturbative, exact set of stochastic differential equations
for the reduced density matrix of the open system, in which the role
of the environment is played by Gaussian stochastic processes in real
and imaginary time which ``replace'' its harmonic oscillators. It
also includes an imaginary time evolution to exactly account for the
chosen thermal equilibrium initial condition (although a broader class
of initial conditions can also be introduced \citep{Grabert1988}),
and allows for a simple and general closed form description of the
evolution of the reduced density matrix. In particular, it is able
to faithfully capture the transient dynamics of driving the system
away from the full system-environment equilibrium caused by any local
(acting only on the open system), and possibly time-dependent, perturbation.
Hence, in this method one starts from the equilibrium density matrtix
of the entire system at a certain temperature $T$, and then considers
exact time evolution of the open system reduced density matrix under
the influence of an arbitrary system-local perturbation switched on
at a particular time. 

While the general proof in Ref. \citep{ourpaper} showed that a reduced
description of the open system is possible, the feasibility of numerical
implementation nor resulting simulations were presented. This paper
represents the first application of the ESLE method to a calculation
\emph{in silico.} Our aim here is (i) to introduce a numerical implementation
of the ESLE; more specifically, we explain how the noises, cross-correlated
with each other, can be generated on a computer, and (ii) apply this
formalism to a spin-boson system driven from equilibrium, with particular
attention paid to how the ESLE modifies the predicted short-time evolution
and how its solution at long times depends on the initial preparation
of the system. 

The rest of this paper will be structured as follows: Sec. II briefly
introduces the ESLC method to provide the essential theoretical basis
needed, which is then applied to a simplified Hamiltonian that enables
us to reduce the ESLE description to only three Gaussian noises. In
Sec. III methods used to generate the noise terms in the ESLE are
discussed. Sec. IV presents the spin-Boson Hamiltonian and results
of the simulations using the ESLE, and we close with a discussion
in Sec. V on the scope of the applications and current limitations
of the ELSE. 

\section{Extended Stochastic Liouville Equation}

\subsection{General theory}

In our model the open system (denoted by a general argument $q$)
is described by an arbitrary Hamiltonian $H_{q}$ (which could be
time-dependent), while the environment is described by a set of harmonic
oscillators (with masses $m_{i}$), with a potential which is quadratic
in their displacement coordinates $\xi=\left\{ \xi_{i}\right\} $.
The interaction between the two subsystems is linear in environment
coordinates $\xi_{i}$, but may involve an arbitrary function of the
open system coordinates, $f_{i}\left(q\right)$. Explicitly, the total
quantum Hamiltonian is given by:

\begin{equation}
H_{\textrm{tot}}(q,\xi,t)=H_{q}(q,t)+\sum_{i}\frac{\zeta_{i}^{2}}{2m_{i}}+\frac{1}{2}\sum_{i,j}\Lambda_{ij}\xi_{i}\xi_{j}-\sum_{i}f_{i}(q)\xi_{i}\label{eq:totalhamiltoniancoordinateform}
\end{equation}
Here $\zeta_{i}$ is the momentum operator conjugate to the displacement
$\xi_{i}$ , and $\Lambda=\left(\Lambda_{ij}\right)$ is the environment
force-constant matrix. This Hamiltonian generalises the CL model in
two ways: firstly, the coupling term is generalised from a bilinear
in CL to a more general one (although still linear in the bath coordinates\footnote{Note that a generalisation to a quadratic coupling (with respect to
the atomic displacements $\xi_{i}$) is also possible, at least formally.}); secondly, the environment model is written via the site representation,
rather than via normal modes, which allows for a more natural extraction
of the parameters of the bath Hamiltonian from the whole system Hamiltonian
(see, e.g., \citep{QGLE-2016}). 

As mentioned in the Introduction, in most treatments considering dynamics
of an open system it is assumed that at initial time $t_{0}$ the
system and environment are \emph{partitioned, }i.e. the full system-environment
density matrix is the tensor product of the open system density matrix
$\rho$ and the environment matrix $\rho_{{\rm env}}$:

\begin{equation}
\rho_{{\rm tot}}\left(t_{0}\right)=\rho\left(t_{0}\right)\otimes\rho_{{\rm env}}\left(t_{0}\right)
\end{equation}
As was demonstrated in our previous work \citep{ourpaper}, instead
of the aforementioned partitioned approximation, the initial state
of the system may be described by a preparation of the canonical density
matrix for the extended system \citep{Grabert1988}:

\begin{equation}
\rho_{{\rm tot}}\left(t_{0}\right)=Q\rho_{\beta}\left(t_{0}\right)Q^{\dagger}
\end{equation}
where
\begin{equation}
\rho_{\beta}\left(t_{0}\right)\equiv\frac{1}{Z_{\beta}}\mathrm{e}^{-\beta H_{0}}\label{eq:Canonical_Density}
\end{equation}
is the initial canonical density matrix, $H_{0}=H_{{\rm tot}}\left(t_{0}\right)$
is the initial Hamiltonian, $\beta=1/k_{B}T$ is the inverse temperature
and $Z_{\beta}=\mbox{Tr}\left(e^{-\beta H_{0}}\right)$ is the corresponding
partition function of the entire system. Here $Q$ is an operator
acting only on the open system. Although this operator may be chosen
in various ways to reflect specific initial conditions, in the present
work we restrict ourselves to the system initially being in full thermal
equilibrium, i.e. we choose $Q=1$. 

With some elementary manipulations, it is possible to derive an exact
evolution of the reduced density matrix $\rho(t)$ describing the
open system. The set of equations governing this evolution constitutes
the ESLE (for full details see \citep{ourpaper}). The ESLE consists
of two stochastic operator evolutions, which, upon averaging over
realisations of the Gaussian stochastic noises it contains (see below),
give the exact open system (i.e. reduced) density matrix at any time
$t$. The first evolution is in imaginary time $\tau$, evolving an
initial density matrix $\overline{\rho}(\tau)$ from $\tau=0$ up
to $\tau=\hbar\beta$, via the following equation:

\begin{equation}
-\hbar\partial_{\tau}\overline{\rho}(\tau)=\left(H_{q}\left(t_{0}\right)-\sum_{i}\bar{\mu}_{i}\left(\tau\right)f_{i}\left(q\right)\right)\overline{\rho}(\tau)\label{eq:ESLE_image_time_many_noises}
\end{equation}
This is then followed by a real time evolution:

\begin{equation}
i\hbar\partial_{t}\tilde{\rho}\left(t\right)=\left[H_{q}\left(t\right),\tilde{\rho}\left(t\right)\right]_{-}-\sum_{i}\left(\eta_{i}\left(t\right)\left[f_{i}\left(q\right),\tilde{\rho}\left(t\right)\right]_{-}+\frac{\hbar}{2}\nu_{i}\left(t\right)\left[f_{i}\left(q\right),\tilde{\rho}\left(t\right)\right]_{+}\right)\label{eq:ESLErealtimemanynoises}
\end{equation}
Here $\left([\ldots]_{+}\right)$ $[\ldots]_{-}$ is the (anti-) commutator.
The two equations are coupled via the condition that the end-point
of the imaginary evolution is the initial condition for the real-time
stochastic dynamics:
\begin{equation}
\left.\overline{\rho}\left(\tau\right)\right|_{\tau=\hbar\beta}=\left.\tilde{\rho}\left(t\right)\right|_{t=0}\label{eq:averaged-Ro}
\end{equation}
To initialise the evolution of $\overline{\rho}(\tau)$, we start
with $\left.\overline{\rho}\left(\tau\right)\right|_{\tau=0}=1$,
evolve it in time and then normalise to unity at the end of the evolution.

The functions $\eta_{i}\left(t\right)$ and $\nu_{i}\left(t\right)$
($\bar{\mu_{i}}\left(\tau\right)$) are zero-mean, complex, Gaussian
noises in real (imaginary) time, and the actual (physical) reduced
density matrix of the open system is obtained by averaging over all
realisations (indicated by $\left\langle ...\right\rangle _{r}$)
of these noise processes:
\begin{equation}
\rho\left(t\right)=\left\langle \tilde{\rho}\left(t\right)\right\rangle _{r}
\end{equation}
The origin of the noise comes from the application of a two-time Hubbard-Stratonovich
transformation \citep{HSTransform} to the influence functional derived
for the environment. This provides an exact mapping of deterministic,
temporally non-local dynamics to a stochastic, local model, provided
the noise correlation functions satisfy the following conditions: 

\begin{equation}
\left\langle \eta_{i}\left(t\right)\eta_{j}\left(t^{\prime}\right)\right\rangle _{r}=\hbar L_{ij}^{R}\left(t-t^{\prime}\right)\label{eq:Corr_F_Eta_Eta-1}
\end{equation}
\begin{equation}
\left\langle \eta_{i}(t)\nu_{j}\left(t^{\prime}\right)\right\rangle _{r}=2i\Theta\left(t-t^{\prime}\right)L_{ij}^{I}\left(t-t^{\prime}\right)\label{eq:Corr_F_Eta_Nu-1}
\end{equation}

\begin{align}
\left\langle \eta_{i}\left(t\right)\bar{\mu}_{j}\left(\tau\right)\right\rangle _{r} & =-\hbar K_{ij}\left(t-i\tau\right)\label{eq:Corr_F_Eta_Mu-1}
\end{align}
\begin{equation}
\left\langle \bar{\mu}_{i}\left(\tau\right)\bar{\mu}_{j}\left(\tau^{\prime}\right)\right\rangle _{r}=\hbar\left[L_{ij}^{e}\left(\tau-\tau^{\prime}\right)-L_{ij}^{o}\left(\left|\tau-\tau^{\prime}\right|\right)\right]\label{eq:Corr_F_Mu_Mu-1}
\end{equation}

\begin{equation}
\left\langle \nu_{i}\left(t\right)\nu_{j}\left(t^{\prime}\right)\right\rangle _{r}=\left\langle \nu_{i}\left(t\right)\bar{\mu}_{j}\left(\tau\right)\right\rangle _{r}=0\label{eq:Corr_F_Zero-1}
\end{equation}
where $\Theta(t)$ is the Heaviside function and the various kernels
are given by:

\begin{equation}
L_{ij}^{R}\left(t\right)=\frac{1}{\sqrt{m_{i}m_{j}}}\sum_{\lambda}\frac{e_{\lambda i}e_{\lambda j}}{2\omega_{\lambda}}\coth\left(\frac{1}{2}\omega_{\lambda}\hbar\beta\right)\cos\left(\omega_{\lambda}\left(t-t^{\prime}\right)\right)\label{eq:Real_L_all_times}
\end{equation}

\begin{equation}
L_{ij}^{I}\left(t\right)=-\frac{1}{\sqrt{m_{i}m_{j}}}\sum_{\lambda}\frac{e_{\lambda i}e_{\lambda j}}{2\omega_{\lambda}}\coth\left(\frac{1}{2}\omega_{\lambda}\hbar\beta\right)\sin\left(\omega_{\lambda}\left(t-t^{\prime}\right)\right)\label{eq:Imag_L_all_times}
\end{equation}

\begin{equation}
L_{ij}^{e}\left(\tau\right)=\frac{1}{\sqrt{m_{i}m_{j}}}\sum_{\lambda}\frac{e_{\lambda i}e_{\lambda j}}{2\omega_{\lambda}}\coth\left(\frac{1}{2}\hbar\beta\omega_{\lambda}\right)\cosh\left(\omega_{\lambda}\tau\right)\label{eq:L_Even}
\end{equation}

\begin{equation}
L_{ij}^{o}\left(\tau\right)=\frac{1}{\sqrt{m_{i}m_{j}}}\sum_{\lambda}\frac{e_{\lambda i}e_{\lambda j}}{2\omega_{\lambda}}\sinh\left(\omega_{\lambda}\tau\right)\label{eq:L_Odd}
\end{equation}
\begin{equation}
K_{ij}\left(t-i\tau\right)=\frac{1}{\sqrt{m_{i}m_{j}}}\sum_{\lambda}\frac{e_{\lambda i}e_{\lambda j}}{2\omega_{\lambda}}\frac{\cosh\left(\omega_{\lambda}\left(\frac{\hbar\beta}{2}-\tau-it\right)\right)}{\sinh\left(\frac{1}{2}\beta\hbar\omega_{\lambda}\right)}\label{eq:Real_K_Complex_Times}
\end{equation}
Here $e_{\lambda}$ are the eigenvectors of the dynamical matrix $\mathbf{D}=\left(D_{ij}\right)$
of the environemnt, where $D_{ij}=\Lambda_{ij}/\sqrt{m_{i}m_{j}}$,
with eigenvalues $\omega_{\lambda}^{2}$. 

Note that for a partitioned initial condition, there is no coupling
between the open system and environment at $t\leq t_{0}$. This approach
is formally equivalent to neglecting the imaginary time evolution
(setting $\tilde{\rho}\left(t_{0}\right)=\rho\left(t_{0}\right)$)
and removing cross-correlations between the real and imaginary time
evolutions (i.e. setting $K_{ij}\left(t-i\tau\right)$ to zero for
all $i$ and $j$). The initial density matrix of the open system,
$\tilde{\rho}\left(t_{0}\right)$, can then be chosen arbitrarily.
This choice may be based on physical insight or alternately as a solution
of the imaginary time evolution, Eq. (\ref{eq:ESLE_image_time_many_noises}).
In the latter case it would correspond to the exact reduced density
matrix for the thermalised extended system.

\subsection{A simplified model }

The prescription considered so far requires three noises ($\eta_{i}(t),$$\nu_{i}(t)$
and $\bar{\mu}_{i}(t)$) per bath lattice site $i$. Alternatively,
one can use the eigenstates of the dynamical matrix $\mathbf{D}$
and normal modes $x_{\lambda}$ instead of the site representation.
Formally we replace $m_{i}\rightarrow1$, $i\rightarrow\lambda$,
$\Lambda_{ij}\rightarrow\omega_{\lambda}^{2}\delta_{\lambda\lambda^{\prime}}$,
$e_{i\lambda^{\prime}}\rightarrow e_{\lambda\lambda^{\prime}}=\delta_{\lambda\lambda^{\prime}}$
and $f_{i}(q)\rightarrow f_{\lambda}(q)$, which reduces the description
to the standard Caldeira-Leggett model \citep{Leggett1987} albeit
with a slightly more general coupling term, $-\sum_{\lambda}f_{\lambda}(q)\xi_{\lambda}$.
In this representation all the correlation function matrices become
diagonal, e.g. 
\[
L_{\lambda\lambda^{\prime}}^{R}(t)=\delta_{\lambda\lambda^{\prime}}L_{\lambda\lambda}^{R}(t)=\delta_{\lambda\lambda^{\prime}}\frac{1}{2\omega_{\lambda}}\coth\left(\frac{1}{2}\omega_{\lambda}\hbar\beta\right)\cos\left(\omega_{\lambda}t\right)
\]
 The main simplification then comes by assuming that, up to a scaling
factor, the $q$ dependences in the coupling functions $f_{\lambda}(q)$
are identical, i.e. $f_{\lambda}(q)=c_{\lambda}f(q)$. In this prescription
it is possible to collectively redefine the noise terms reducing them
to just three distinct terms. Taking the $\eta_{i}\rightarrow\eta_{\lambda}$
noises in Eq. (\ref{eq:ESLErealtimemanynoises}) as an example:

\begin{equation}
\sum_{i}\eta_{i}\left(t\right)\left[f_{i}\left(q\right),\tilde{\rho}\left(t\right)\right]_{-}\Rightarrow\sum_{\lambda}\eta_{\lambda}\left(t\right)\left[f_{\lambda}\left(q\right),\tilde{\rho}\left(t\right)\right]_{-}=\sum_{\lambda}c_{\lambda}\eta_{\lambda}\left(t\right)\left[f\left(q\right),\tilde{\rho}\left(t\right)\right]_{-}=\eta\left(t\right)\left[f\left(q\right),\tilde{\rho}\left(t\right)\right]_{-}
\end{equation}
with $\eta(t)=\sum_{\lambda}c_{\lambda}\eta_{\lambda}(t)$. The correlation
function of this combined noise will be given by:

\begin{equation}
\left\langle \eta\left(t\right)\eta\left(t^{\prime}\right)\right\rangle _{r}=\hbar\sum_{\lambda\lambda^{\prime}}c_{\lambda}c_{\lambda^{\prime}}L_{\lambda\lambda^{\prime}}^{R}\left(t\right)=\hbar\sum_{\lambda}c_{\lambda}^{2}L_{\lambda\lambda}^{R}\left(t\right)=\hbar\sum_{\lambda}\frac{c_{\lambda}^{2}}{2\omega_{\lambda}}\coth\left(\frac{1}{2}\omega_{\lambda}\hbar\beta\right)\cos\left(\omega_{\lambda}\left(t-t^{\prime}\right)\right)
\end{equation}
In the continuum limit we can replace the sum over bath modes with
an integration over frequency:

\begin{equation}
\sum_{\lambda}\frac{c_{\lambda}^{2}}{2\omega_{\lambda}}\ldots\Rightarrow\int_{0}^{\infty}\frac{{\rm d}\omega}{\pi}\,\left[\pi\sum_{\lambda}\frac{c_{\lambda}^{2}}{2\omega_{\lambda}}\delta\left(\omega-\omega_{\lambda}\right)\right]\ldots=\int_{0}^{\infty}\frac{{\rm d}\omega}{\pi}\,I\left(\omega\right)\ldots
\end{equation}
Here $I\left(\omega\right)$ is the bath spectral density, which is
formally chosen depending on the specific model which couples oscillators
of the environment and the system. In this paper we shall use an Ohmic
spectral density, given by:

\begin{equation}
I\left(\omega\right)=\alpha\omega\left[1+\left(\frac{\omega}{\omega_{c}}\right)^{2}\right]^{-2}\label{eq:Ohmic}
\end{equation}
where $\omega_{c}$ is some cut-off frequency. The parameter $\alpha$
is proportional to squares of the $c_{\lambda}$ coefficients and
hence describes the effective bath coupling strength.

Similarly to the $\eta(t)$ noise, two other collective noises, $\nu(t)$
and $\bar{\mu}(\tau)$, are introduced, giving three noises in total.
In the reduced case, which we shall use in the rest of the paper,
the ESLE can be completely described by two stochastic differential
equations, one for the initialisation in the imaginary time,

\begin{equation}
-\hbar\partial_{\tau}\overline{\rho}(\tau)=\left[H_{q}\left(t_{0}\right)-\bar{\mu}\left(\tau\right)f\left(q\right)\right]\overline{\rho}(\tau)\label{eq:ESLEimagtime-1}
\end{equation}
and another for the propagation in real time:

\begin{equation}
i\hbar\partial_{t}\tilde{\rho}\left(t\right)=\left[H_{q}\left(t\right),\tilde{\rho}\left(t\right)\right]_{-}-\left\{ \eta\left(t\right)\left[f\left(q\right),\tilde{\rho}\left(t\right)\right]_{-}+\frac{\hbar}{2}\nu\left(t\right)\left[f\left(q\right),\tilde{\rho}\left(t\right)\right]_{+}\right\} \label{eq:ESLErealtime-1}
\end{equation}
where the corresponding correlation functions are given by the following
equations:

\begin{equation}
\left\langle \eta\left(t\right)\eta\left(t^{\prime}\right)\right\rangle _{r}=\hbar\int_{0}^{\infty}\frac{{\rm d}\omega}{\pi}\,I\left(\omega\right)\coth\left(\frac{1}{2}\omega\hbar\beta\right)\cos\left(\omega\left(t-t^{\prime}\right)\right)\equiv K_{\eta\eta}\left(t-t^{\prime}\right)\label{eq:Corr_F_Eta_Eta}
\end{equation}
\begin{equation}
\left\langle \eta(t)\nu\left(t^{\prime}\right)\right\rangle _{r}=-2i\Theta\left(t-t^{\prime}\right)\int_{0}^{\infty}\frac{{\rm d}\omega}{\pi}\,I\left(\omega\right)\sin\left(\omega\left(t-t^{\prime}\right)\right)\equiv K_{\eta\nu}\left(t-t^{\prime}\right)\label{eq:Corr_F_Eta_Nu}
\end{equation}

\begin{align}
\left\langle \eta\left(t\right)\bar{\mu}\left(\tau\right)\right\rangle _{r} & =-\hbar\int_{0}^{\infty}\frac{{\rm d}\omega}{\pi}\,I\left(\omega\right)\frac{\cosh\left(\omega\left(\frac{\hbar\beta}{2}-\tau-it\right)\right)}{\sinh\left(\frac{1}{2}\beta\hbar\omega\right)}\equiv K_{\eta\bar{\mu}}\left(t-i\tau\right)\label{eq:Corr_F_Eta_Mu}
\end{align}
\begin{equation}
\left\langle \bar{\mu}\left(\tau\right)\bar{\mu}\left(\tau^{\prime}\right)\right\rangle _{r}=\hbar\int_{0}^{\infty}\frac{{\rm d}\omega}{\pi}\,I\left(\omega\right)\left(\cosh\left(\omega\left(\tau-\tau^{\prime}\right)\right)\coth\left(\frac{1}{2}\omega\hbar\beta\right)-\sinh\left(\omega\left(\tau-\tau^{\prime}\right)\right)\right)\equiv K_{\bar{\mu}\bar{\mu}}\left(\tau-\tau^{\prime}\right)\label{eq:Corr_F_Mu_Mu}
\end{equation}

\begin{equation}
\left\langle \nu\left(t\right)\nu\left(t^{\prime}\right)\right\rangle _{r}=\left\langle \nu\left(t\right)\bar{\mu}\left(\tau\right)\right\rangle _{r}=0\label{eq:Corr_F_Zero}
\end{equation}
While this set of equations represents an exact description, it requires
the generation of noises in two different time dimensions, which obey
the specific statistical relationships detailed above. The numerical
implementation of these noises will now be discussed.

\section{Generating Noises}

\begin{figure}[H]
\begin{centering}
\includegraphics[height=6cm]{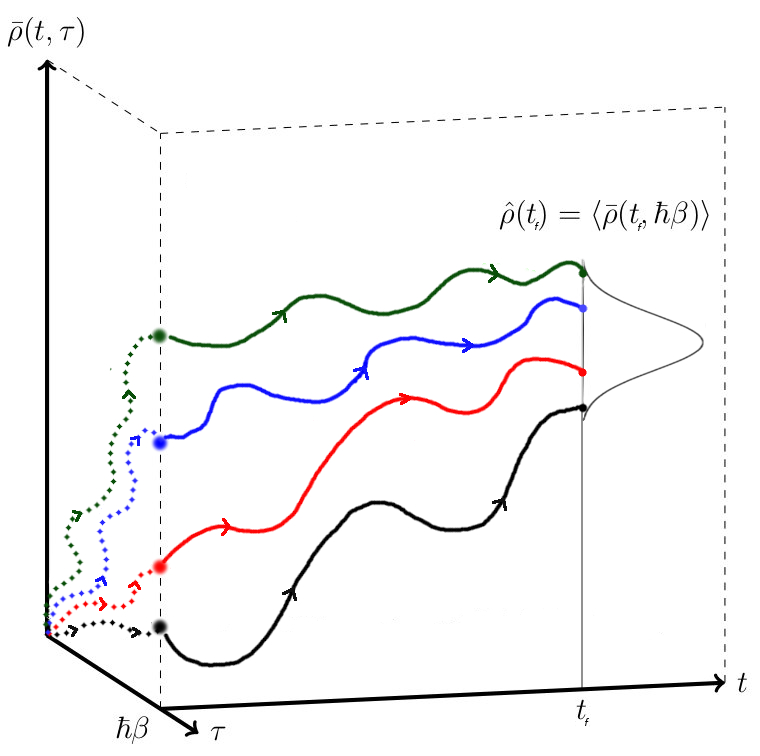}
\par\end{centering}
\caption{Representative trajectories for the evolution of the system. First
there is an evolution in imaginary time up to $\tau=\beta\hbar$ (dashed
lines), before evolving in real time (solid lines) from this point
up to time $t_{f}$. Different colours correspond to different realisations
of the noises. The average of the final points gives the physical
reduced density matrix at that time (indicated at time $t_{f}$).
Reproduced with permission from Ref. \citep{ourpaper}.\label{fig:matrix_evolution_in_2_time_dimensions}}
\end{figure}

The numerical scheme for the ESLE is in principle rather simple, illustrated
by Fig. \ref{fig:matrix_evolution_in_2_time_dimensions}. The proceduce
is to generate a realisation of the noises satisfying Eqs. (\ref{eq:Corr_F_Eta_Eta})-(\ref{eq:Corr_F_Zero}),
evolve the density matrix according to Eqs. (\ref{eq:ESLEimagtime-1})
and (\ref{eq:ESLErealtime-1}), and finally average over realisations,
Eq. (\ref{eq:averaged-Ro}). In this section we shall discuss how
the three noises are generated; the method used can easily be generalised
to any number of noises. 

The general scheme to generate coloured Gaussian noises is well known
\citep{PhysRevE.91.032125}, and the situation is only slightly complicated
by the existence of ESLE's non-stationary cross-time correlations.
We begin by splitting each noise into sub-terms which are only correlated
with one other term across the noises: 

\begin{equation}
\eta\left(t\right)=\eta_{\eta}\left(t\right)+\eta_{\nu}\left(t\right)+\eta_{\bar{\mu}}\left(t\right)
\end{equation}

\begin{equation}
\nu\left(t\right)=\nu_{\eta}\left(t\right)
\end{equation}

\begin{equation}
\bar{\mu}\left(t\right)=\bar{\mu}_{\bar{\mu}}\left(\tau\right)+\bar{\mu}_{\eta}\left(\tau\right)
\end{equation}
For instance, the $\eta_{\eta}$ noise is the auto-correlative part
of the total $\eta$ noise, and has a non-zero correlation only with
itself, while the $\eta_{\bar{\mu}}$ noise only correlates with the
$\bar{\mu}_{\eta}$ term. To enforce this effective ``noise-orthogonality'',
we express each term as a convolution between a filtering kernel (which
are denoted as $G_{\eta\eta}$, $G_{\eta\nu}$, etc.) and one of a
number of real white noise processes $\left\{ x_{i}\left(t\right),\bar{x}_{i}(\tau)\right\} $,
which have the property:

\begin{equation}
\left\langle x_{i}\left(t\right)x_{j}\left(t^{\prime}\right)\right\rangle _{r}=\delta_{ij}\delta\left(t-t^{\prime}\right)\label{eq:orthonorm-1}
\end{equation}

\begin{equation}
\left\langle \bar{x}_{i}\left(\tau\right)\bar{x}_{j}\left(\tau^{\prime}\right)\right\rangle _{r}=\delta_{ij}\delta\left(\tau-\tau^{\prime}\right)\label{eq:orthonorm-2}
\end{equation}

\begin{equation}
\left\langle x_{i}\left(t\right)\bar{x}_{j}\left(\tau\right)\right\rangle _{r}=0\label{eq:orthonorm-3}
\end{equation}

Given that the time (real or imaginary) is simply a parameter in the
noise process, the distinction between $x_{i}$ and $\bar{x}_{j}$
is one of notational convenience, rather than an expression of any
fundamentally dissimilar statistics. The various components of the
three complex noises we need are generated by the following convolutions
of the filtering kernels with the white noises:

\begin{equation}
\eta_{\eta}\left(t\right)=\int_{-\infty}^{\infty}{\rm d}t\ G_{\eta\eta}\left(t-t^{\prime}\right)x_{1}(t^{\prime})\label{eq:etaetacomponent}
\end{equation}

\begin{equation}
\eta_{\nu}\left(t\right)=\int_{-\infty}^{\infty}{\rm d}t^{\prime}\ G_{\eta\nu}\left(t-t^{\prime}\right)\left(x_{2}(t^{\prime})+ix_{3}\left(t^{\prime}\right)\right)\label{eq:etanucomponent}
\end{equation}

\begin{equation}
\nu_{\eta}\left(t\right)=\int_{-\infty}^{\infty}{\rm d}t^{\prime}\ G_{\nu\eta}\left(t-t^{\prime}\right)\left(x_{3}(t^{\prime})+ix_{2}\left(t^{\prime}\right)\right)\label{eq:nuetacomponent}
\end{equation}

\begin{equation}
\bar{\mu}_{\bar{\mu}}\left(\tau\right)=\int_{-\beta\hbar}^{\beta\hbar}{\rm d}\tau^{\prime}\ G_{\bar{\mu}\bar{\mu}}\left(\tau-\tau^{\prime}\right)\bar{x}_{1}\left(\tau^{\prime}\right)\label{eq:mumucomponent}
\end{equation}
\begin{equation}
\eta_{\bar{\mu}}\left(t\right)=\int_{0}^{\beta\hbar}{\rm d}\tau^{\prime}\ G_{\eta\bar{\mu}}\left(t,\tau^{\prime}\right)\left(\bar{x}_{2}(\tau^{\prime})+i\bar{x}_{3}\left(\tau^{\prime}\right)\right)\label{eq:etamucomponent}
\end{equation}

\begin{equation}
\bar{\mu}_{\eta}\left(\tau\right)=\int_{0}^{\beta\hbar}{\rm d}\tau^{\prime}\ G_{\bar{\mu}\eta}\left(\tau,\tau^{\prime}\right)\left(\bar{x}_{3}(\tau^{\prime})+i\bar{x}_{2}\left(\tau^{\prime}\right)\right)\label{eq:muetacomponent}
\end{equation}
Here the limits on integrations over imaginary time reflect the fact
that $\tau$ is constrained to lie within the interval $[0,\beta\hbar]$.
In this construction the various filtering kernels are yet to be determined.
Importantly, as the only physically relevant constraints on the noises
are the physical kernels, this linear filtering ansatz will be valid
provided we can establish a consistent mapping between the physical
$K$ (see Eqs. (\ref{eq:Corr_F_Eta_Eta}) - (\ref{eq:Corr_F_Zero}))
and filtering $G$ kernels. Note that in all cases we assume the filtering
kernels $G$ have the same stationarity properties (i.e. they depend
only on time differences) as the corresponding physical kernels $K$.
Hence the\emph{ cross-time correlation} filtering kernels $G_{\eta\bar{\mu}}\left(t,\tau\right)$
and $G_{\bar{\mu}\eta}\left(\tau,\tau^{\prime}\right)$, that are responsible for the cross-time correlation $\left\langle \eta\left(t\right)\bar{\mu}\left(\tau\right)\right\rangle _{r}$ between real and imaginary times noises (see Eq. ([eq:newcrosstimekernelident]) later on) involves both real and imaginary times, are not assumed to be stationary, unlike the other filtering kernels.

It can easily be seen that, with the above choice, $\left\langle \eta\left(t\right)\eta\left(t^{\prime}\right)\right\rangle _{r}=\left\langle \eta_{\eta}\left(t\right)\eta_{\eta}\left(t^{\prime}\right)\right\rangle _{r}$,
$\left\langle \nu\left(t\right)\eta\left(t^{\prime}\right)\right\rangle _{r}=\left\langle \nu_{\eta}\left(t\right)\eta_{\nu}\left(t^{\prime}\right)\right\rangle _{r}$,
$\left\langle \bar{\mu}\left(\tau\right)\bar{\mu}\left(\tau^{\prime}\right)\right\rangle _{r}=\left\langle \bar{\mu}_{\bar{\mu}}\left(\tau\right)\bar{\mu}_{\bar{\mu}}\left(\tau^{\prime}\right)\right\rangle _{r}$,
and $\left\langle \eta\left(t\right)\bar{\mu}\left(\tau\right)\right\rangle _{r}=\left\langle \eta_{\bar{\mu}}\left(t\right)\bar{\mu}_{\eta}\left(\tau\right)\right\rangle _{r}$;
all other correlation functions are identically equal to zero because
of the design imposed ``orthonormality'' of the white noises, Eqs.
(\ref{eq:orthonorm-1})-(\ref{eq:orthonorm-3}). For instance, 
\[
\left\langle \nu\left(t\right)\nu\left(t^{\prime}\right)\right\rangle _{r}=\int_{-\infty}^{\infty}\int_{-\infty}^{\infty}{\rm d}t_{1}{\rm d}t_{2}\ \left[G_{\nu\eta}\left(t-t_{1}\right)G_{\nu\eta}\left(t^{\prime}-t_{2}\right)\left(\left\langle x_{3}\left(t_{1}\right)x_{3}\left(t_{2}\right)\right\rangle _{r}\right.\right.
\]

\begin{equation}
\left.\left.-\left\langle x_{2}\left(t_{1}\right)x_{2}\left(t_{2}\right)\right\rangle _{r}+2i\left\langle x_{3}\left(t_{1}\right)x_{2}\left(t_{2}\right)\right\rangle _{r}\right)\right]=0
\end{equation}
as by Eq. (\ref{eq:Corr_F_Zero}). 

To find the correspondence between the physical and filtering kernels,
we substitute the assumed functional form of each noise given above
into their relations (\ref{eq:Corr_F_Eta_Eta}) - (\ref{eq:Corr_F_Zero})
with the physical kernels. Explicit evaluation of this auto-correlative
component yields:

\[
\left\langle \eta\left(t\right)\eta\left(t^{\prime}\right)\right\rangle _{r}=\int_{-\infty}^{\infty}\int_{-\infty}^{\infty}{\rm d}t_{1}{\rm d}t_{2}\ G_{\eta\eta}\left(t-t_{1}\right)G_{\eta\eta}\left(t^{\prime}-t_{2}\right)\left\langle x_{1}\left(t_{1}\right)x_{1}\left(t_{2}\right)\right\rangle 
\]

\begin{equation}
=\int_{-\infty}^{\infty}{\rm d}t_{1}\ G_{\eta\eta}\left(t-t_{1}\right)G_{\eta\eta}\left(t^{\prime}-t_{1}\right)
\end{equation}
leading to the first of the kernel mappings:

\begin{equation}
K_{\eta\eta}\left(t-t^{\prime}\right)=\int_{-\infty}^{\infty}{\rm d}t_{1}\ G_{\eta\eta}\left(t-t_{1}\right)G_{\eta\eta}\left(t^{\prime}-t_{1}\right)\label{eq:onenoiseconvolution}
\end{equation}
The physical kernel is therefore expressible as a self-convolution
of the filtering kernel, and can be further simplified using its Fourier
representation:

\begin{equation}
\tilde{K}_{\eta\eta}\left(\omega\right)=\left|\tilde{G}_{\eta\eta}\left(\omega\right)\right|^{2}\label{eq:etaetafourierkernel}
\end{equation}
This equation is the \emph{only} constraint on $G_{\eta\eta}$, and
any solution to this equation yields a valid filtering kernel. In
this particular case the solution is unique (up to a phase), but we
shall see later that cross-correlative mappings do \emph{not} uniquely
define the relevant filtering kernels, and hence some choice exists
which can be exploited. Given that the physical kernel here is both
real and symmetric, we may constrain the filtering kernel to have
the same properties and express it simply as:

\begin{equation}
\tilde{G}_{\eta\eta}\left(\omega\right)=\sqrt{\tilde{K}_{\eta\eta}\left(\omega\right)}
\end{equation}

The auto-correlative component of the $\bar{\mu}$ noise has the same
properties as above, provided we extend the integration domain of
Eq. (\ref{eq:mumucomponent}) and periodically extend $G_{\bar{\mu}\bar{\mu}}\left(\tau-\tau^{\prime}\right)$
across this domain. Then the filtering kernel has an identical mapping
in Fourier space:

\begin{equation}
\tilde{G}_{\bar{\mu}\bar{\mu}}\left(\omega\right)=\sqrt{\tilde{K}_{\bar{\mu}\bar{\mu}}\left(\omega\right)}
\end{equation}

There are also two non-zero cross-correlative terms to consider: the
real time correlation between the $\eta$ and $\nu$ noises, and the\lyxdeleted{Mccaul}{Mon Dec 18 12:52:56 2017}{
} cross-time correlation between $\eta$ and $\bar{\mu}$. In the
first case, we have:

\begin{equation}
\left\langle \eta\left(t\right)\nu\left(t^{\prime}\right)\right\rangle _{r}=K_{\eta\nu}\left(t-t^{\prime}\right)=2i\int_{-\infty}^{\infty}{\rm d}t_{1}\ G_{\eta\nu}\left(t-t_{1}\right)G_{\nu\eta}\left(t^{\prime}-t_{1}\right)\label{eq:realcrosscorreqn}
\end{equation}
 or in Fourier space:

\begin{equation}
\tilde{K}_{\eta\nu}\left(\omega\right)=2i\tilde{G}_{\eta\nu}\left(\omega\right)\tilde{G}_{\nu\eta}^{*}\left(\omega\right)
\end{equation}
Unlike with the auto-correlative processes, we are left with a degree
of choice in the form of the two filtering kernels. Here we take the
simple expedient of choosing one of the kernels as a delta function,
$G_{\nu\eta}\left(t\right)=\delta\left(t\right)$ or $\tilde{G}_{\nu\eta}\left(\omega\right)=1$.
The second filtering kernel may therefore be straightforwardly identified
as:

\begin{equation}
\tilde{G}_{\eta\nu}\left(\omega\right)=-\frac{i}{2}\tilde{K}_{\eta\nu}\left(\omega\right)
\end{equation}

We now turn our attention to the final cross correlation, for which
we obtain:

\begin{equation}
\left\langle \eta\left(t\right)\bar{\mu}\left(\tau\right)\right\rangle _{r}=K_{\eta\bar{\mu}}\left(t-i\tau\right)=2i\int_{0}^{\beta\hbar}{\rm d}\tau^{\prime}\ G_{\eta\bar{\mu}}\left(t,\tau^{\prime}\right)G_{\bar{\mu}\eta}\left(\tau,\tau^{\prime}\right)\label{eq:newcrosstimekernelident}
\end{equation}
In this case we cannot use the Fourier transformation to simplify
the expression of the filtering kernels, as the physical kernel is
inherently non-stationary. Once again it is convenient to set one
kernel as a delta function, $G_{\bar{\mu}\eta}\left(\tau,\tau^{\prime}\right)=\delta\left(\tau-\tau^{\prime}\right)$,
giving the form of the other kernel as:

\begin{equation}
K_{\eta\bar{\mu}}\left(t-i\tau\right)=2iG_{\eta\bar{\mu}}\left(t,\tau\right)
\end{equation}
which completes the mapping between the physical and filtering kernels.  Note that setting either of the filtering kernels $G_{\eta\bar{\mu}}\left(t,\tau\right)$ or $G_{\bar{\mu}\eta}\left(\tau,\tau^{\prime}\right)$ to zero results in the loss of correlations between the imaginary and real time evolutions, which essentially corresponds to the reduced SLE method with the initial density matrix obtained independently from the imaginary time evolution.

Armed with this mapping, the noises are straightforwardly generated
by the convolution of filtering kernels with vectors of white noise
(with variances appropriate to the discretisation of the timestep).
Further details on the generation of these noises, as well as the
numerical solution for the ESLE may be found in the Appendix.

\begin{figure}[p]
\includegraphics[height=5cm]{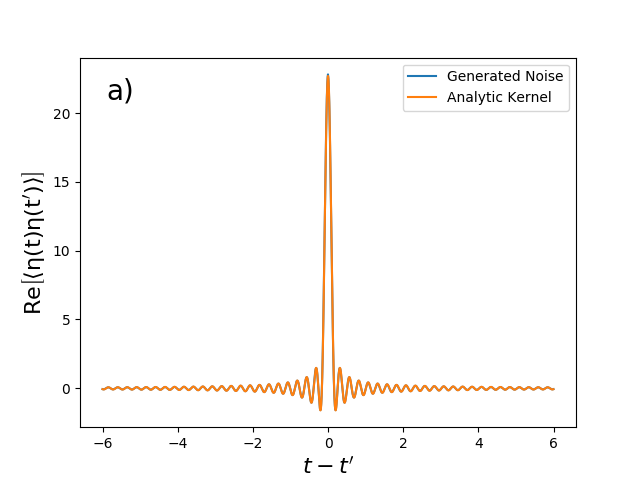}\includegraphics[height=5cm]{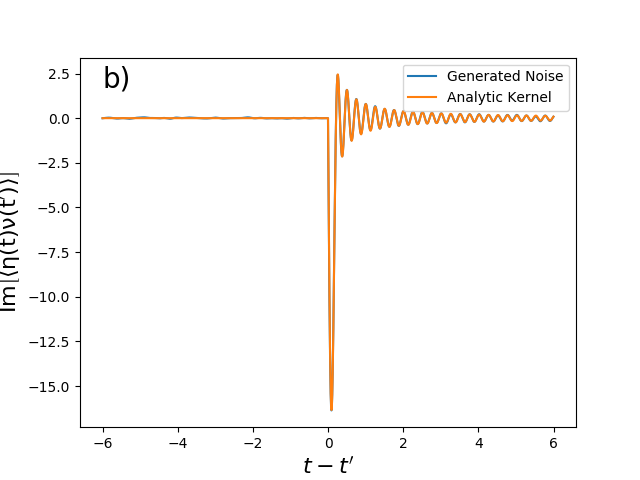}

\includegraphics[height=5cm]{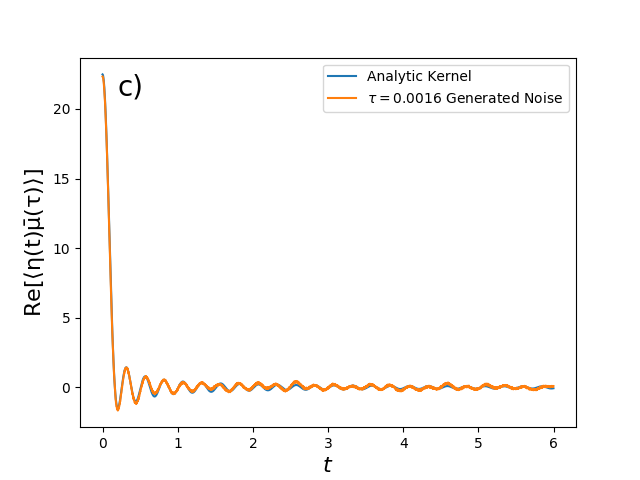}\includegraphics[height=5cm]{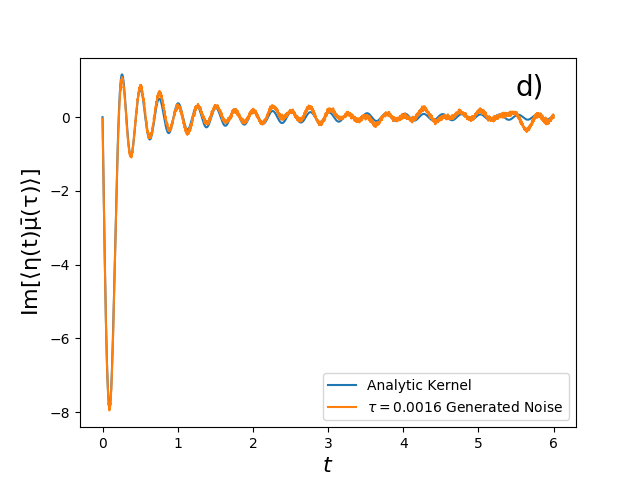}

\includegraphics[height=5cm]{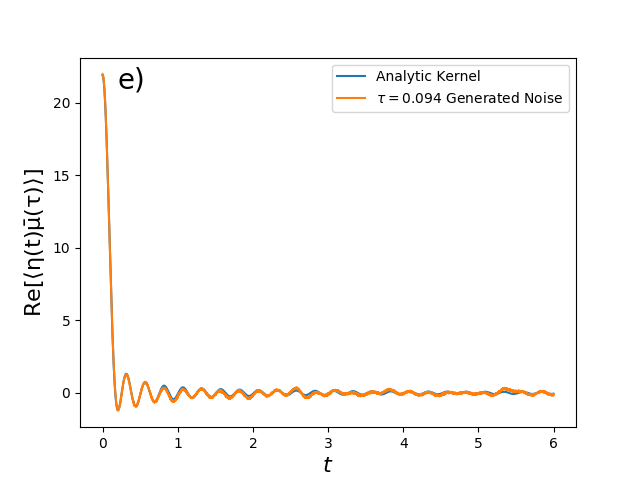}\includegraphics[height=5cm]{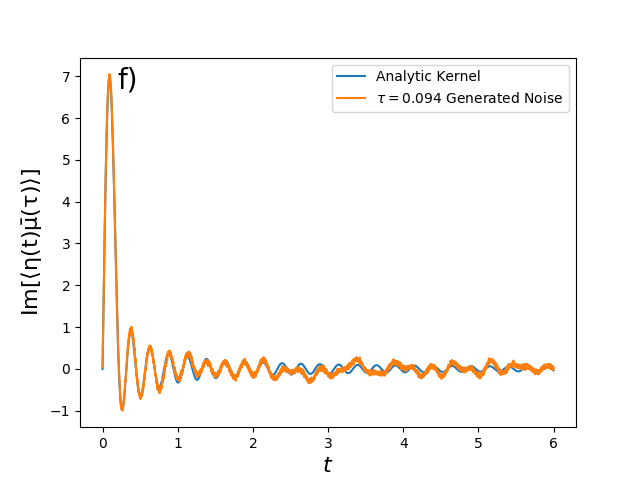}

\includegraphics[height=5cm]{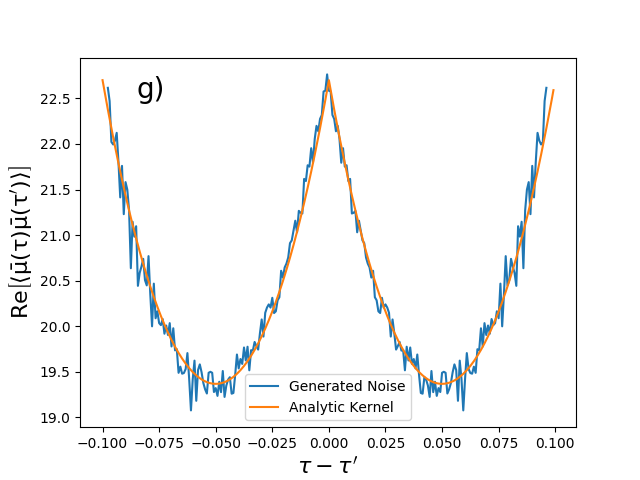}\includegraphics[height=5cm]{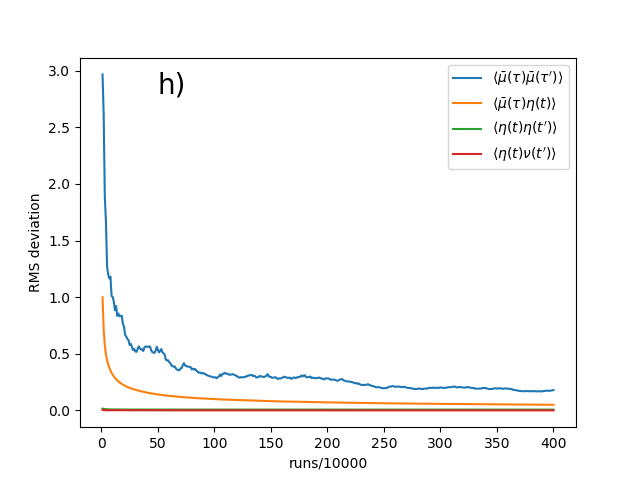}

\caption{Typical correlation functions obtained from generated noise after
$4\times10^{6}$ runs with parameters $\beta=0.1\Delta^{-1}$, $\alpha=0.2$,
$\omega_{c}=25$, using the Ohmic spectral density of Eq. (\ref{eq:Ohmic}).
Cross-time correlations c\textbf{), }d\textbf{) }and e\textbf{), }f\textbf{)}\textbf{\emph{
}}are illustrated with two seperate $\tau$ slices, as the full two-time
correlation forms a 2D surface. Correlations expected to be zero are
not shown, but the \emph{maximum} value found across any of these
functions was on the order of $10^{-4}$, and therefore ignorable.
The final figure shows the RMS deviation between the (real part) of
the generated noise covariances and their respective kernels with
increasing numbers of runs.\label{fig:Typical-Correlation-functions} }
\end{figure}

Typical results for noise generation are shown in Fig. \ref{fig:Typical-Correlation-functions},
demonstrating that excellent convergence to the physical kernels can
be achieved, provided that a sufficient sampling is taken. In particular,
the apparently noisier behaviour of $\left\langle \bar{\mu}\left(\tau\right)\bar{\mu}\left(\tau^{\prime}\right)\right\rangle _{r}$
is due to its relatively small range of values, and the fact that
its cross-time correlative part must be generated with the cruder
direct summation (see Appendix). Panel (f) shows the RMS deviation
for the non-zero correlations as functions of the number of runs used
to generate them; one can see that the overall error is sharply decreased
with the number of runs.

\section{Application to a Driven Spin-Boson Model}

\subsection{Spin-Boson Models}

Consider a two-state system described by the (matrix) Hamiltonian:

\begin{equation}
H_{q}\left(t\right)=\epsilon\left(t\right)\sigma_{z}+\Delta\left(t\right)\sigma_{x}\label{eq:LZ-Hamiltonian}
\end{equation}
Here $\epsilon\left(t\right)$ describes the bias between states,
while $\Delta\left(t\right)$ controls tunnelling between them; $\sigma_{z}$
and $\sigma_{x}$ are the usual Pauli spin-matrices. While using our
formalism it would be possible to consider any spin-boson model, we
shall focus here on two seperate protocols: (i) an equilibrium real
time evolution when the bias $\epsilon(t)$ is kept constant at the
value used to thermalise the whole system during imaginary time evolution,
and (ii) a Landau-Zener type sweep, where the system has been thermalised
at some time in the past, $t_{0}<0$, and then evolved with $\epsilon(t)=\kappa t$.
In both cases the tunneling $\Delta$ is kept constant. Rather than
using an analytic result in the second case (none exists), the asymptotic
limit will be extrapolated numerically from the initial state calculated
by the ESLE. Given that the ESLE evolves from an explicit thermal
state at finite temperature and there is no analytic expression available,
including the one to describe its asymptotic behaviour, the two-state
solution provides a useful numerical benchmark to evaluate the impact
of the environment. 

Coupling this two-state system to an environment of the standard CL
type and using $f_{\lambda}=c_{\lambda}\sigma_{z}$, the total Hamiltonian
reads as:

\begin{equation}
H_{\textrm{tot}}\left(t\right)=\epsilon(t)\sigma_{z}+\Delta\sigma_{x}+\sum_{\lambda}\hbar\omega_{\lambda}b_{\lambda}^{\dagger}b_{\lambda}-\sigma_{z}\sum_{\lambda}c_{\lambda}\left(b_{\lambda}+b_{\lambda}^{\dagger}\right)\label{eq:full-LZ-Hamiltonian}
\end{equation}
This is simply the matrix form of the total Hamiltonian given in Eq.
(\ref{eq:totalhamiltoniancoordinateform}), with the appropriate model-specific
substitutions and the second quantisation for the environment oscillators,
where $b_{\lambda}$ ($b_{\lambda}^{\dagger}$) is the bosonic annhilation
(creation) operator. The last term corresponds to the system-environment
coupling which is proportional to the normal mode displacements of
the environment. The only explicit $t$ dependence in the total Hamiltonian
is contained in the (possible) bias field for the open system. 

To apply the ESLE to this system, we assume that the total system-environment
is allowed initially (at time $t_{0}$) to thermalise having the Hamiltonian
$H_{0}=H_{{\rm tot}}\left(t_{0}\right)$ corresponding to some initial
value of the bias. Explicitly:

\begin{equation}
\rho_{{\rm tot}}\left(t_{0}\right)=\frac{1}{Z_{\beta}}\mathrm{e}^{-\beta H_{0}}
\end{equation}
This initial condition implies the following ESLE equations in imaginary
($0\leq\tau\leq\beta\hbar$) and real ($t\geq t_{0}$) times:

\begin{equation}
-\hbar\partial_{\tau}\overline{\rho}(\tau)=\left[\epsilon\left(t_{0}\right)\sigma_{z}+\Delta\sigma_{x}-\bar{\mu}\left(\tau\right)\sigma_{z}\right]\overline{\rho}(\tau)\label{eq:ESLEspinbosonimagtime}
\end{equation}

\begin{equation}
i\hbar\partial_{t}\tilde{\rho}\left(t\right)=\left[\epsilon\left(t\right)\sigma_{z}+\Delta\sigma_{x},\tilde{\rho}\left(t\right)\right]_{-}-\eta\left(t\right)\left[\sigma_{z},\tilde{\rho}\left(t\right)\right]_{-}-\frac{\hbar}{2}\nu\left(t\right)\left[\sigma_{z},\tilde{\rho}\left(t\right)\right]_{+}\label{eq:ESLEspinbosonrealtime}
\end{equation}
In other words, we consider the initial condition to be parametrised
by $t_{0}$ with real-time dynamics either keeping that value of the
bias (the first protocol) or linearly driving the system away from
its thermal state (the second). 

Finally, we should be precise in our definition of strong-coupling.
This is usually measured by the parameter $\alpha$ in the spectral
density, Eq. (\ref{eq:Ohmic}). It has been shown that for $\alpha<\frac{1}{2}$
there is a coherent evolution, but crossing through the point $\alpha=\frac{1}{2}$
causes a phase change to incoherent spin dynamics \citep{Leggett1987,HUR20082208}.
Beyond this at $\alpha>1$ the system enters a localised regime where
tunnelling between the two states is completely suppressed (formally
the bath coupling renormalises the tunelling element to $\Delta\to0$
). These behaviours are peculiar to the spin-boson model, and not
indicative of a general restriction of the parameter space the ESLE
is capable of simulating exactly. Our results will focus on the regime
$\alpha<\frac{1}{2}$, where we should expect coherent, damped oscillations
in the spin expectations. 

\subsection{Equilibrium}

As a sanity check, we first test the ESLE for a time-independent bias.
Given the ESLE is thermalised in imaginary time exactly, we expect
the real-time evolution to show no change in the density matrix. 

\begin{figure}[h]
\begin{centering}
\includegraphics[height=6cm]{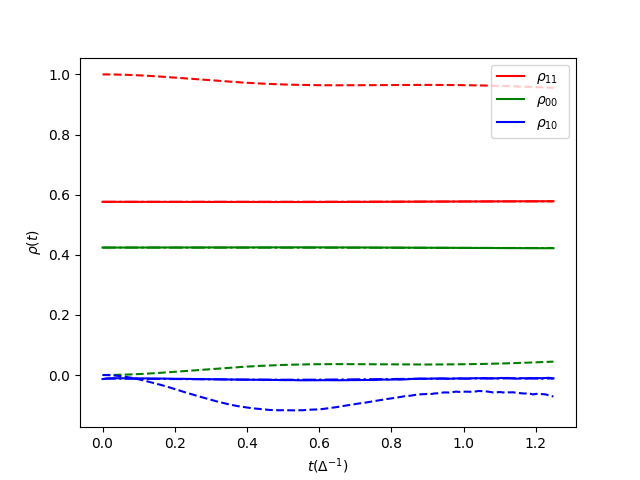}
\par\end{centering}
\caption{ESLE evolution of $\rho$ for a time-independent system, averaged
over $1\times10^{5}$ runs (solid lines). Dashed lines indicate an
equivalent SLE evolution from $\rho_{ij}=\delta_{i1}\delta_{j1}$,
while dash-dotted lines (that practically coincide with the solid
ones) are SLE evolutions from $\rho_{ij}$=$\left\langle \bar{\rho}_{ij}\left(\hbar\beta\right)\right\rangle _{r}$.
Here $\Delta=1$, $\epsilon=5\Delta$, $\beta=0.1\Delta^{-1}$ $\alpha=0.05$,
and $\omega_{c}=200$. While the ESLE shows small fluctuations, the
density matrix remains on average constant. The SLE evolution from
$\rho_{ij}=\delta_{i1}\delta_{j1}$ predicts a relaxation to the ESLE
values, but on a timescale not accessible by the simulation.\label{fig:eqmflatline}}
\end{figure}

Figure \ref{fig:eqmflatline} shows the density matrix components
of both the ESLE, and a comparative simulation running the real time
part of the ESLE without the cross-time correlations. This reduced
case corresponds to the Stochastic Liouville Equation (SLE) \citep{Stockburger2004}
based on the partitioned approach. In the SLE we initialise the density
matrix from two initial conditions: (i) $\rho_{ij}=\delta_{i1}\delta_{j1}$
and (ii) the density matrix predicted by the ESLE imaginary time evolution,
$\rho_{ij}$=$\left\langle \bar{\rho}_{ij}\left(\hbar\beta\right)\right\rangle _{r}$. 

For both the ESLE and the SLE simulation initialised from the imaginary-time
evolution end-point, small oscillations in the components are observed,
but they remain on average constant. This demonstrates that the cross-time
correlations in the ESLE have little to no effect at equilibrium.
This is as expected: any cross-time correlations in the noise rapidly
die out as the system evolves in real time, and the noises evolving
the SLE and ESLE become statistically identical. If the SLE simulation
(started from the ESLE initial condition) evolved differently to the
ESLE, then the ESLE would echo that behaviour later in time- a manifestly
unphysical scenario at equilibrium. 

The SLE simulation initialised at $\rho_{ij}=\delta_{i1}\delta_{j1}$
shows a relaxation of the spin. It is expected that this relaxation
will converge to the same steady state as predicted by the other simulations,
but it does so on a timescale not fully accessible by our numerical
scheme. This is due to the fact that in equilibrium, the cut-off frequency
of the bath spectrum must be sufficiently large that any energy the
spin system dissipates to the bath is returned in a finite time. If
this is not the case thermal equilibrium is not possible, as the bath
acts as an energy sink (causing the spin to relax). At the same time,
from a numerical perpective, higher cut-off frequencies require a
smaller time-step to avoid non-physical resonances. This is doubly
problematic, as the timescale for the SLE to relax increases with
cut-off frequency \citep{bistablesystem}, while the stochastic simulation
itself is step-limited. That is, numerical instabilities at longer
times require excessive averaging to eliminate, i.e. many more simulations
are needed for sampling, which is extremely demanding computationally.
At lower cut-off frequencies, all simulations are observed to converge
to the same state, but the ESLE displays an unphysical spin-relaxation
from the thermalised state due to the low cut-off. 

\subsection{Landau-Zener Protocol}

Here we shall consider fully non-equilibrium simulations, in which
the bias is linearly driven from the value $\epsilon\left(t_{0}\right)=\kappa t_{0}$
used for the equilibration (imaginary time evolution). This spin-boson
model is particularly useful in this case, as for specific initial
conditions the asymptotic behaviour can be analytically derived. In
the zero temperature case, when the system is started in the state
\begin{equation}
\rho\left(-\infty\right)=\rho_{0}^{LZ}=\left(\begin{array}{cc}
1 & 0\\
0 & 0
\end{array}\right)
\end{equation}
and is decoupled from the environment oscillators, the asymptotic
survival probability of that state is given by:

\begin{equation}
P_{LZ}=\exp\left(-\frac{\pi\Delta^{2}}{\hbar\kappa}\right)\label{eq:LZ-limit}
\end{equation}

This protocol is known as the Landau-Zener (LZ) sweep, with $P_{LZ}$
first derived by Zener by recasting the system as a Weber equation
\citep{Zener696}. The result may also be found via contour integration
\citep{Wittig2005} or direct evaluation of the time-ordered propagator
\citep{Rojo2010}. This protocol has numerous experimental realisations,
for example in Rydberg atoms \citep{PhysRevA.23.3107} or Bose-Einstein
condensates \citep{PhysRevA.90.013616}. It has also been proven that
the survival probability for the state is the same even with a $\sigma_{z}$
coupling to a dissipative environment (of the Caldeira-Leggett type),
with the caveat that the total system must be prepared in the ground
state at zero temperature and evolved from $t_{0}\to-\infty$ \citep{PhysRevLett.97.200404,PhysRevB.75.214308}.
Furthermore, numerical evaluations using a Stochastic Schr{\"o}dinger
Equation (SSE) have shown that even in the case where the evolution
is started from some finite time sufficiently far in the past (for
a fixed initial spin), provided the bath coupling is sufficiently
weak $\left(\alpha<0.2\right)$, $P_{LZ}$ is still recovered at zero
temperature \citep{Orth2013}. At stronger couplings however deviations
from $P_{LZ}$ in the asymptotic state were observed, confirming that
generally even at zero temperature the asymptotic spin state in a
dissipative system depends on both the bath coupling strength and
the initial preparation of the system. 

As our theory enables to provide an \emph{exact dynamics} of the spin-boson
system density matrix, it is interesting to explore the validity of
the LZ sweep limit (\ref{eq:LZ-limit}) in detail, both at finite
temperature and with the envoronment coupling.

\subsubsection{General time behaviour}

To model the LZ type sweep, we thermalise the system at some time
$t_{0}<0$ in the past with the bias $\epsilon_{0}=\kappa t_{0}$
and then evolve it in real time for $t>t_{0}$. For the purposes of
achieving a quicker relaxation to the asymptotic state, the cut-off
frequency $\omega_{c}=25$ was chosen for all simulations. Unlike
in the equilibrium case, where a large cut-off frequency was used
to ensure the energy scale of the bath was always much greater than
that of the spin, in the driven case in our simulations the system
starts and ends with considerably stronger bias. As all the dynamical
changes occur in a window around $t=0$, the effect of reducing the
cut-off frequency is to narrow the region where the state transitions
may happen. In addition, we set $\Delta=1$, which means that effectively
all parameters of the system are scaled to units of $\Delta$. 

\begin{figure}[h]
\begin{centering}
\includegraphics[height=6cm]{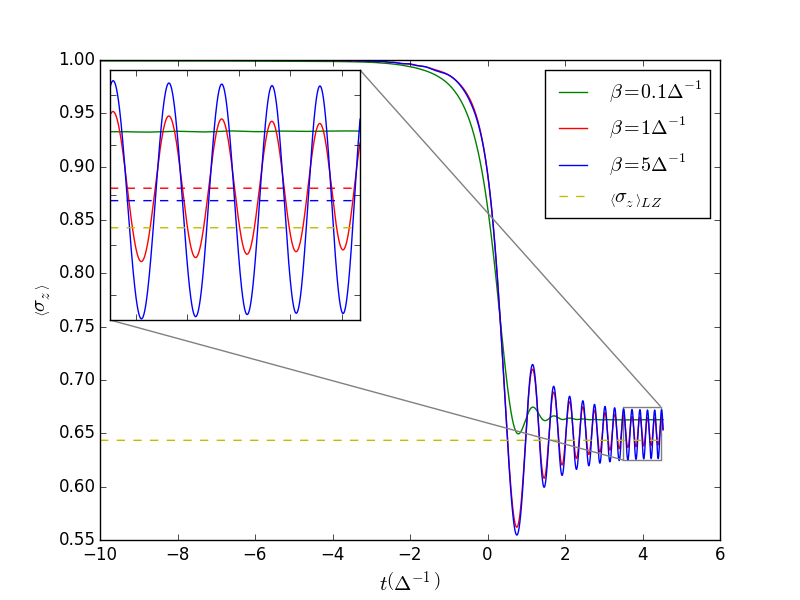}
\par\end{centering}
\caption{ESLE evolution for a fast sweep $t_{0}=-10\Delta^{-1}$, $\kappa=8\Delta^{2}$,
and $\alpha=0.05$, sampled over $1\times10^{6}$ runs. Red and blue
dashed lines indicate extrapolated asymptotes for the two lowest temperature
simulations. We observe that even at weak coupling the asymptotic
value of $\left\langle \sigma_{z}\right\rangle $ deviates from the
LZ expectation $\left\langle \sigma_{z}\right\rangle _{LZ}$, although
lower temperature asymptotic states lie closer to the LZ limit.\label{fig:longtimeLZ}}
\end{figure}

Fig. \ref{fig:longtimeLZ} shows an ESLE evolution of $\left\langle \sigma_{z}\right\rangle =\text{Tr}\left(\rho(t)\sigma_{z}\right)=\rho_{11}(t)-\rho_{22}(t)$
at finite temperature, where parameters were chosen such that the
initial density matrix approaches that of the LZ initial condition,
$\rho\left(t_{0}\right)\approx\rho_{0}^{LZ}$, although the evolution
still begins from a finite time in the past. We expect that from this
initial condition the cross-time correlations (which rapidly attenuate
with time) are suppressed when evolving from a regime where the bias
field is initially much stronger than thermal effects. This limiting
case therefore also serves as a check that the ESLE predicts evolutions
consistent with partitioned methods.

One can see that at finite temperatures the asymptotic behaviour does
not necessarily converge to the LZ result even at weak coupling. In
addition, while the mean state of the system rapidly converges to
its asymptotic limit, the amplitude of oscillations around this state,
and their rate of decay appears dependent on temperature, with oscillation
amplitude decaying slower at lower temperatures. We also observe that
the mean value approaches the LZ value as the temperature is lowered.
This can be explained by the final state (and its convergence to the
LZ limit) being dependent on the size of the temporal region where
the bias field is comparable to the strength of thermal fluctuations.
This region, where $\left|\beta\kappa t\right|\lessapprox1$ is when
thermal effects will have the greatest impact on the dynamics of the
system, as elsewhere the bias field dominates the system evolution.
Therefore, we should expect the asymptotic state at lower temperature
to lie closer to the LZ limit. Unfortunately, the time required for
oscillations to decay sufficiently to confirm this is much longer
at lower temperature. Given the excessive computational cost of longer
simulation times in the ESLE (see Section \ref{sec:discussion}),
in Fig. \ref{fig:longtimeLZ} the asymptotic states for the two lowest
temperatures are extrapolated from (oscillating) data. From this we
conclude that high temperatures (or slow sweeps) allow thermal effects
to increase the asymptotic $\left\langle \sigma_{z}\right\rangle $
value away from $\left\langle \sigma_{z}\right\rangle _{LZ}$, consistent
with earlier SSE results \citep{Orth2010}. 

\subsubsection{Coupling-Strength Dependence}

\begin{figure}[h]
\begin{centering}
\includegraphics[height=5cm]{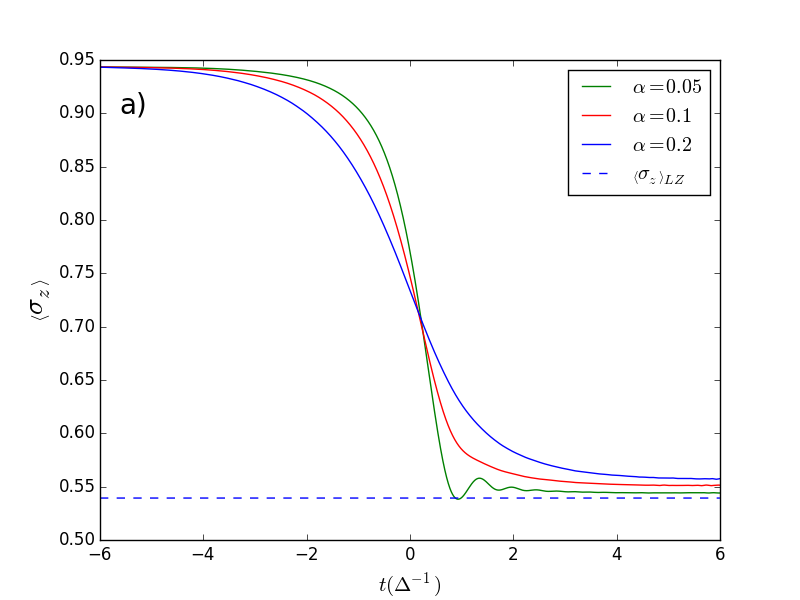}\includegraphics[height=5cm]{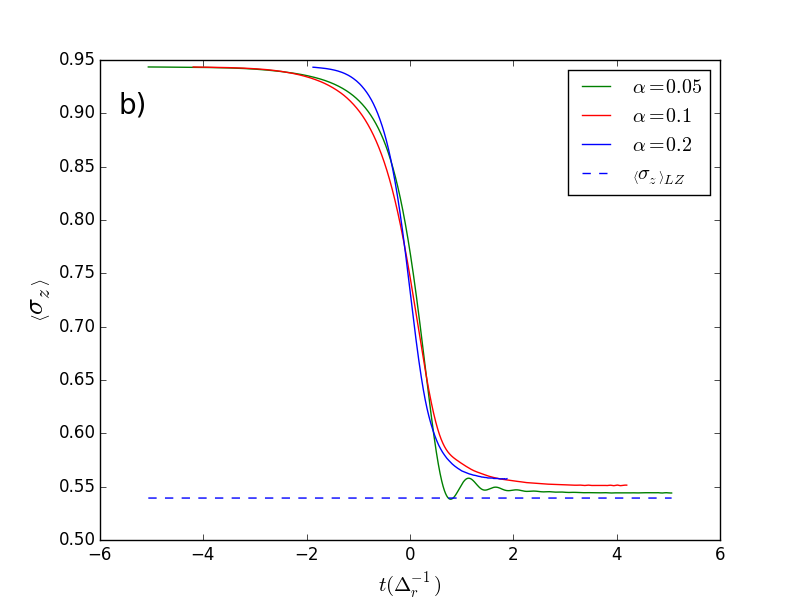}
\par\end{centering}
\caption{\textbf{(}a)\textbf{ }Real time $\left\langle \sigma_{z}\right\rangle $
dynamics for the system with parameters $t_{0}=-6\Delta^{-1}$, $\kappa=6\Delta^{2}$,
and $\beta=0.1\Delta^{-1}$, after sampling with $1\times10^{6}$
runs. \textbf{(}b) The same dynamics are rescaled such that the curves
of different $\alpha$ lie nearly on top of one another, demonstrating
the spin-bath coupling renormalising the tunnelling element of the
two level system. \label{fig:alphavariable}}
\end{figure}

Fig. \ref{fig:alphavariable}\textbf{ (}a\textbf{) }shows the real
time dynamics of $\left\langle \sigma_{z}\right\rangle $ for the
LZ sweep at different coupling strengths. Here simulations are started
from a sufficiently small $t_{0}$ such that the calculated initial
density matrices $\tilde{\rho}\left(t_{0}\right)$ are distinguishable
from $\rho_{0}^{LZ}$. Comparing results in Fig. \ref{fig:alphavariable},
we find that the bath coupling has two principal effects. First, oscillatory
behaviour in the spin expectations is suppressed by increasing bath
coupling, as expected. The asymptotic survival probability also increases,
for the same reason as when increasing temperature. Indeed, stronger
coupling allows thermal effects to have a stronger influence on the
evolution lifting the $\left\langle \sigma_{z}\right\rangle $ asymptote.
This phenomenon can also be understood as the system-bath coupling
renormalising the characteristic frequency scale of oscillations,
resulting in a quicker thermalisation. This scaling can be expressed
in terms of the renormalised tunelling element \citep{Leggett1987,PhysRevA.93.032119}:

\begin{equation}
\Delta_{r}=\Delta\left(\frac{\Delta}{\omega_{c}}\right)^{\frac{\alpha}{1-\alpha}}
\end{equation}
Given this scaling is an argument based purely on renormalising the
system Hamiltonian, we should expect it to hold regardless of the
initial condition chosen. It should be noted here that the ESLE is
at root a faithful representation for a particular kind of initial
condition, and should not affect the dynamical properties of a system.
Fig. \ref{fig:alphavariable} \textbf{(}b\textbf{) }shows the $\left\langle \sigma_{z}\right\rangle $
dynamics when time is scaled via this renormalised tunnelling element,
demonstrating that the curves of varying $\alpha$ scale on top of
each other, hence serving as another consistency check that the ESLE
produces physically reasonable results. 

\subsubsection{Comparison to Partitioned Evolution}

\begin{figure}[H]
\begin{centering}
\includegraphics[height=5cm]{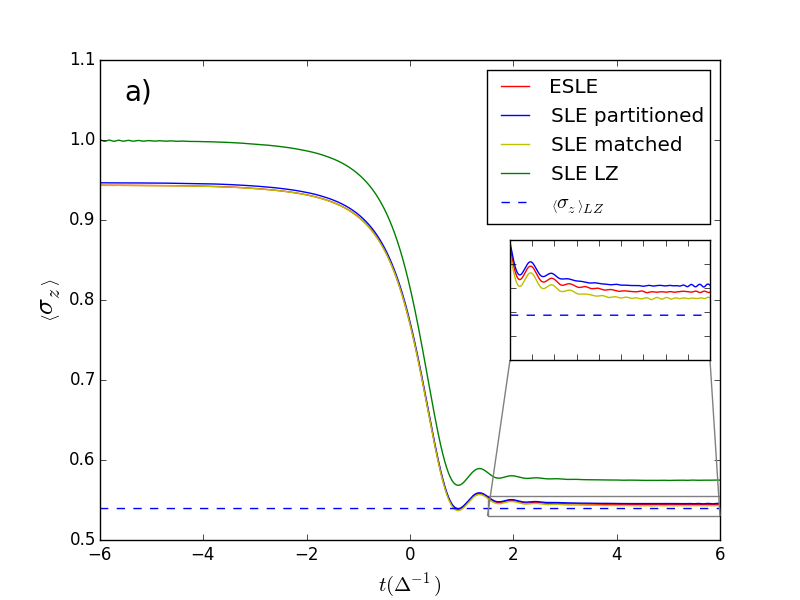}\includegraphics[height=5cm]{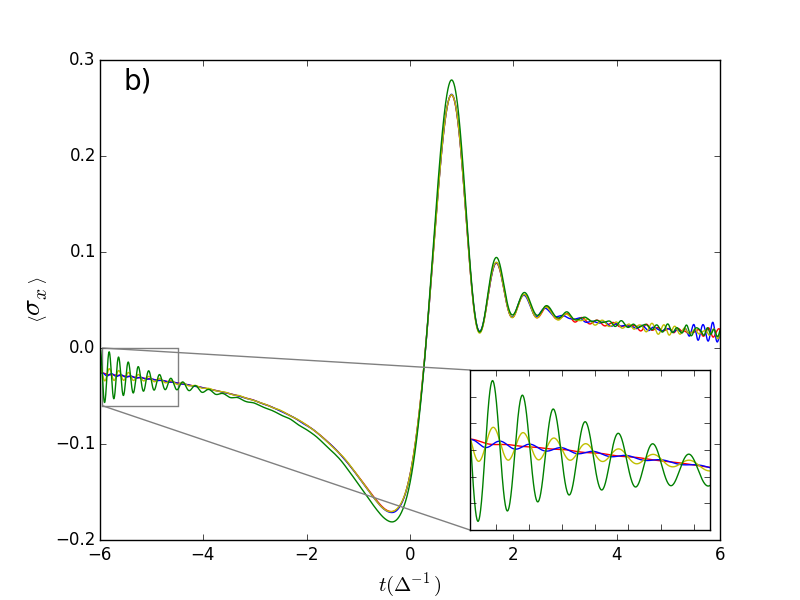}
\par\end{centering}
\begin{centering}
\includegraphics[height=5cm]{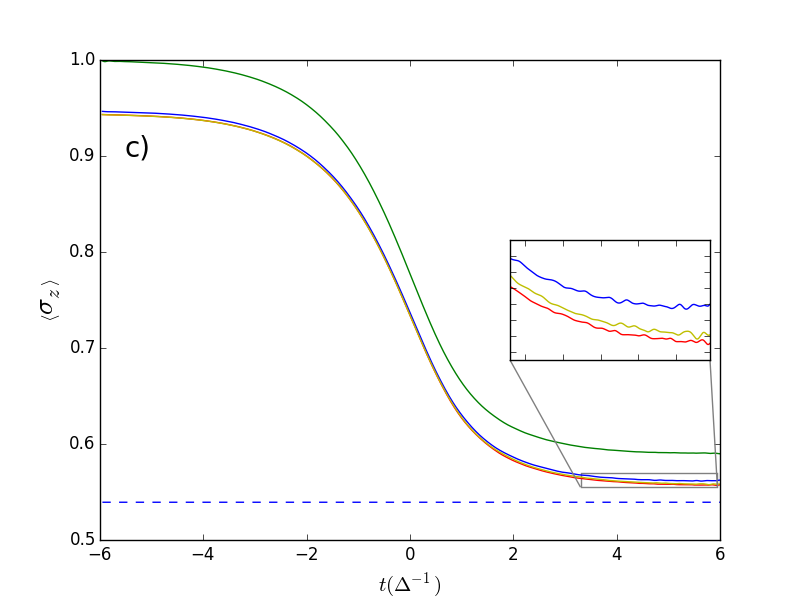}\includegraphics[height=5cm]{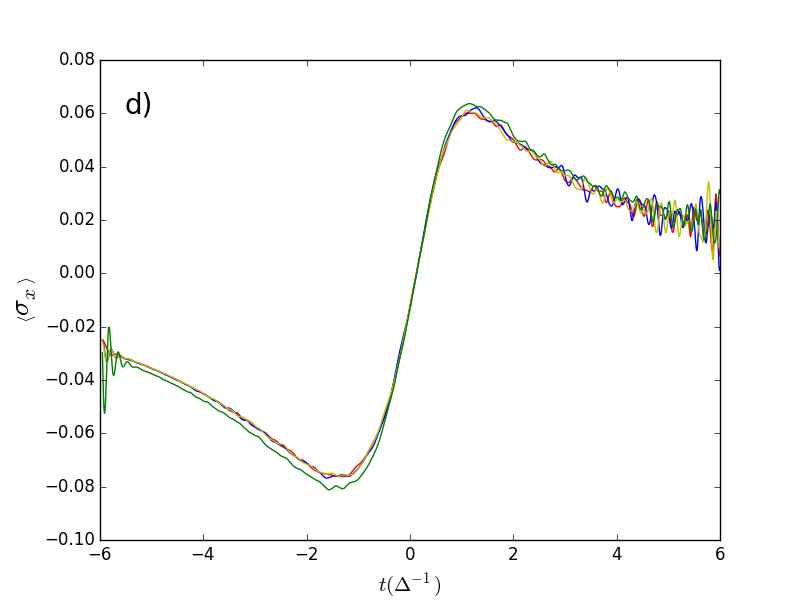}
\par\end{centering}
\caption{Real time spin dynamics for the system with parameters $t_{0}=-6\Delta^{-1}$,
$\kappa=6\Delta^{2}$, $\beta=0.1\Delta^{-1}$, sampled over $6\times10^{5}$
runs. \textbf{(}a\textbf{)} and \textbf{(}b\textbf{)} show $\left\langle \sigma_{z}\right\rangle $
and $\left\langle \sigma_{x}\right\rangle $, respectively, at weak
coupling $\alpha=0.05$, while \textbf{(}c\textbf{)} and (d\textbf{)
}give\textbf{ }$\left\langle \sigma_{z}\right\rangle $ and $\left\langle \sigma_{x}\right\rangle $
for a stronger coupling $\alpha=0.2$. \label{fig:alphasmall}}
\end{figure}

We now compare the full ESLE to an SLE evolution purely in real time.
Using the SLE, we may consider three different initial conditions: 
\begin{itemize}
\item ``SLE LZ'', where the system is evolved from the Landau-Zener initial
condition: $\tilde{\rho}\left(t_{0}\right)=\rho_{0}^{LZ}$;
\item ``SLE matched'', where the initial condition is that calculated from
the averaged ESLE imaginary time evolution, i.e. from the exact reduced
density matrix, $\tilde{\rho}\left(t_{0}\right)=\left\langle \bar{\rho}\left(\hbar\beta\right)\right\rangle _{r}=\frac{1}{Z_{\beta}}{\rm Tr}_{{\rm env}}\left[{\rm \exp\left(-\beta H_{{\rm 0}}\right)}\right]$,
obtained by solving Eq. (\ref{eq:ESLEspinbosonimagtime}); this evolution
differs from the exact ESLE only in that the cross-time correlation
is set to zero;
\item ``SLE partitioned'', where the partitioned approximation is made
to the initial state: $\tilde{\rho}\left(t_{0}\right)=Z^{-1}\exp\left[-\beta H_{q}\left(t_{0}\right)\right]$.
\end{itemize}
Fig. \ref{fig:alphasmall} shows the ESLE solution compared to these
three cases of the SLE at both (a,b) weak and (c,d) strong coupling.
There are several points to note here. The first is that there is
a small (but visible) difference between the initial condition calculated
by the ESLE and the naive initial condition used by the``SLE partitoned''
approach, particularly at strong coupling. All partitioned evolutions
exhibit initial oscillations (which are damped by stronger coupling),
particularly in the coherence $\left\langle \sigma_{x}\right\rangle =\rho_{21}(t)+\rho_{12}(t)$.
Its asymptotic state is also different to the ESLE at both coupling
strengths. The partitioned simulation also displays greater numerical
instability, particularly at strong coupling.

The SLE LZ partitioned simulation differs most from the ESLE in its
transient dynamics. This is consistent with the ESLE's cross-time
correlations, which are expected to have the largest effect at the
start of the real-time dynamics due to the decay of the corresponding
correlation kernels (\ref{eq:Corr_F_Eta_Mu}) at longer times. To
test whether the oscillations observed in this simulation are simply
due to initial conditions, or if the cross-time correlations in the
ESLE actually suppress these oscillations, we compare to the ``SLE
matched'' initial preparation, which starts from the calculated ESLE
initial density matirx but then neglects the cross-time correlations
in its real time evolution.. Comparing this simulation to the ESLE
we find cross-time correlations are responsible for damping small
transient oscillations in the coherence (see \ref{fig:alphasmall}(b)).
From this we conclude that the main contributary factor in the observed
difference between the ESLE and ``SLE partitioned'' is due the choice
of initial condition. 

Examining the long-time behaviour reveals a small gap between the
asymptotic states of the ESLE and ``SLE matched'' that can only be
due to the cross-time correlations missing in the SLE simulation.
This indicates that the cross-time correlations have an observable
effect not only on the transient dynamics, but on the asymptotic state
as well. This is not surprising, given that the Landau-Zener steady
state is known to depend on initial conditions as representing a continuing
non-equilibrium evolution with time-dependent Hamiltonian.

Finally we note that the ``SLE LZ'' simulation gives an asymptotic
value for $\left\langle \sigma_{z}\right\rangle $ significantly higher
than the LZ limit. This is to be expected, given the high temperature
simulated, and the relatively short time in the past the evolution
is begun from. 

\section{discussion\label{sec:discussion}}

In this paper we have presented a numerical application of the exact
ESLE to a driven spin-boson model.

While there are no analytic predictions for evolutions from the exact
initial density matrix presented, the numerical solution for the spin-boson
system dynamics considered here using our exact partitionless method
have been found to be extremely important as it can serve as a reference
when comparing with previous approximate calculations based on the
partitioned approach. In the latter method, cross-correlations between
system preparation (imaginary time evolution or thermalisation) and
real time evolution are artificially missing. In this proof of concept
for the method we have restricted ourselves to relatively short evolutions
at high sweep speeds, to achieve quicker convergence of the results.

We have shown that for a simple system-bath coupling considered here
only three Gaussian noises need to be generated: one for the imaginary
time evolution that brings the entire system (the open system and
the bath) to thermal equilibrium (initial preparation), and two functions
for the real time evolution. The method presented here enabled us
to generate these noises in such a way that all correlation functions
are reproduced. We find, however, that small errors require very large
sampling set, i.e. up to and over $10^{5}$ simulation runs are required
to produce physically reasonable results. 

As a sanity check of the method and its implementation, we first considered
a real time evolution with a cons4tant system Hamiltonian. One would
expect that the real time evolution with the exact density matrix
obtained after thermalisation in imaginary time should remain unchanged,
and this was indeed found to be the case: we have seen that the ESLE
predicts no change from initial conditions in its real time evolution
(as expected). SLE simulations from various initial conditions show
relaxation, but on a timescale where we cannot reliably ascertain
their steady state. Nevertheless, in the long time limit cross-time
correlations die away, and both the ESLE and SLE will be evolved using
noises with identical statistical properties, therefore we should
expect the SLE simulation to converge to the ESLE result. At the same
time, we find that great care is needed in achieving numerically acceptable
results. This stems from the fact that the ESLE is a stochastic differential
equation with multiplicative noise, as well as a first order finite
difference approximation?. This inevitably leads to a limit on the
number of steps it may efficiently simulate before numerical instabilities
dominate. In the future a more sophisticated numerical implementation
\citep{NAKAJIMA1997983,PhysRevE.72.067701} will help results to converge
with a larger timestep and hence allow access to longer timescales.

In addition to the equilibrium simulation, the non-equilibrium problem
of a Landau-Zener sweep (in which the bias in the open system Hamiltonian
is linearly driven) was also simulated. The exact ESLE simulation
is compared specifically with the approximate (partitioned) SLE approach
in which only real time evolution is considered from a chosen initial
density matrix. We observe significant differences between partitioned
evolutions and the ESLE, particularly at stronger coupling. We have
found that the asymptotic behaviour of the ESLE in a Landau-Zener
sweep protocol is \emph{qualitatively} consistent with earlier results,
showing that decreasing temperature and coupling strength brings the
asymptotic solution for the survival probability closer to the known
zero-temperature and zero-coupling result. At larger temperatures
and coupling strengths the asymptotic state deviates significantly
due to the presence of the bath, regardless of the initial condition
used. In particular, even if one chooses to calculate the initial
reduced density matrix exactly by thermalising the whole system, the
cross-time correlations of the ESLE have a small (but observable)
effect in both the transient dynamics \emph{and} the asymptotic state
as compared to the SLE approach where this correlation is switched
off. 

These results highlight behaviours that may prove important in practical
applications, particularly where quantum coherence is a resource,
as there is a small (but persistent) difference between the ESLE and
SLE predictions for driving away from equilibrium, particularly at
short times. In addition, this approximation-free behaviour will affect
the efficiency of quantum heat engines. It has already been shown
that strong bath coupling diminishes performance \citep{PhysRevE.95.032139},
and the results of ESLE calculations show the state of the system
when equilibriated with a heat reservoir is \emph{not} the naive canonical
state.

Finally, the arena of future applications of the ESLE is particularly
broad. As well as the aforementioned heat engines, properties such
as thermal transport and entropy production through a spin system
may be examined with the ESLE, as well as applications to the time
evolution of more complex multi-state open systems interacting with
bosonic fields (phonons and/or photons). These applications would
require generalising the method presented here for generating noises
as more than three noises would be required. To achieve these aims,
a more efficient algorithm for generating noises may also be necessary. 

Concluding, we stress that an important application of the ESLE is
that it could serve as a test bed for verifying approximate analytical
and numerical approaches. The essence of ELSE is that only for a particular
manifestation of the stochastic fields an analytical representation
of the reduced density matrix (and hence a precise form of the corresponding
effective Liouville equation describing a non-unitary evolution) is
possible. Many such evolutions must be sampled in order to get the
final result which is physically meaningful. Hence, the ELSE method
is ultimately a numerical technique, but upon convergence it is capable
of obtaining an \emph{exact }result. The fact that only numerical
results are possible is not necessarily a disadvantage: even though
an analytical solution with this method is out of reach, the fact
that this method does provide an exact (albeit numerical) result is
still very important: firstly, one can obtain exact solutions for
a particular Hamiltonian, and, secondly, the method can also be used
to verify various approximate analytical theories.

\section*{Acknowledgement}

GM is supported by the EPSRC Centre for Doctoral Training in Cross-Disciplinary
Approaches to Non-Equilibrium Systems (CANES, EP/L015854/1).

\section*{Appendix: Numerical Implementation Algorithm}

The process for simulating a single trajectory for the ESLE consists
of two parts: generating noise vectors, and using those in a stochastic
differential equation. When generating noises, we have seen that each
noise is a sum of different components with constructed correlations.
In general the form of these noise components when discretised is:

\begin{equation}
y_{i}=\delta_{t/\tau}\sum_{j}g_{ij}x_{j}\label{eq:discretenoise}
\end{equation}
where $x$ is a complex sum of unit variance white noises scaled by
$\frac{1}{\sqrt{\delta_{t/\tau}}}$ (to give the discrete delta function
correlations), $g_{ij}$ is the (discretised) appropriate filtering
kernel and $\delta_{t}$ (or $\delta_{\tau}$) is the step for the
real (imaginary) time being summed over. The second index on $g$
is necessary for describing the cross-time correlative components
of the $\bar{\mu}$ and $\eta$ noises, where the filtering kernel
is inherently two-dimensional. In this case Eq. (\ref{eq:discretenoise})
must be implemented through direct matrix multiplication. For example,
$\eta_{\bar{\mu}}$ is calculated as:

\begin{equation}
\eta_{\bar{\mu}}\left(t_{i}\right)=\delta_{\tau}\sum_{j=0}^{M}G_{\eta\bar{\mu}}\left(t_{i}\right)\left(\bar{x}_{2}\left(\tau_{j}\right)+i\bar{x}_{3}\left(\tau_{j}\right)\right)
\end{equation}
where $i=1,\ldots,N$, with $N$ being the number of real timesteps
($M$ imaginary timesteps). Fig \ref{fig:Typical-Correlation-functions}
shows this method produces noises with the correct properties, it
is inefficient since it requires $N\times M$ operations to generate
the cross-time component of a noise vector. 

For components of the noise with stationary correlations (i.e. no
mixing of real and imaginary time), the filtering kernel matrix is
expressible as a vector of time differences, $g_{ij}\to g_{i-j}$.
It is therefore much more efficient to use the Fast Fourier Transform
(FFT) to convolve the filtering kernels with the white noises. The
FFT uses the Discrete Fourier Transform (DFT), which transforms a
length $N$ sequence in the following manner:

\begin{equation}
{\rm DFT}_{\alpha}\left[\vec{y}\right]=Y_{\alpha}=\sum_{j=0}^{N-1}y_{j}\exp\left(-2\pi i\frac{\alpha j}{N}\right)
\end{equation}
with the inverse:

\begin{equation}
{\rm DFT}_{j}^{-1}\left[\vec{Y}\right]=y_{j}=\frac{1}{N}\sum_{\alpha=0}^{N-1}Y_{\alpha}\exp\left(2\pi i\frac{\alpha j}{N}\right)
\end{equation}
The DFT has a \emph{circular} convolution theorem:

\begin{equation}
\sum_{k=0}^{N-1}\left(g^{(N)}\right)_{j-k}x_{k}={\rm DFT}_{j}^{-1}\left[\sum_{\alpha}G_{\alpha}X_{\alpha}\right]
\end{equation}
where $g^{(N)}$ is the periodic extension of the sequence $g$:

\begin{equation}
\left(g^{(N)}\right)_{i}\equiv g_{i({\rm mod}N)}
\end{equation}
It is possible to obtain the linear convolution from the circular
convolution (and hence the DFT convolution theorem) by padding both
the filtering kernel and noise vectors with a large number of zeros
\citep{DFTconvolution}. This allows an efficient generation of stationary
noise components as:

\begin{equation}
y_{j}=\delta_{t/\tau}{\rm DFT}_{j}^{-1}\left[\sum_{\alpha}{\rm DFT}_{\alpha}\left[\vec{g}\right]{\rm DFT}_{\alpha}\left[\vec{x}\right]\right]
\end{equation}
When a noise vector has been generated, the density matrix is evolved
by the first order discretisation:

\begin{equation}
\overline{\rho}_{\tau+\delta_{\tau}}=\overline{\rho}_{\tau}+\delta_{\tau}\left[\epsilon\sigma_{z}+\Delta\sigma_{x}-\bar{\mu}_{\tau}\sigma_{z}\right]\overline{\rho}_{\tau}\label{eq:ESLEspinbosonimagtime-1}
\end{equation}

\begin{equation}
\tilde{\rho}_{t+\delta_{t}}=\tilde{\rho}_{t}+\delta_{t}\left\{ \left[\epsilon\sigma_{z}+\Delta\sigma_{x},\tilde{\rho}_{t}\right]_{-}-\eta_{t}\left[\sigma_{z},\tilde{\rho}_{t}\right]_{-}-\frac{\hbar}{2}\nu_{t}\left[\sigma_{z},\tilde{\rho}_{t}\right]_{+}\right\} \label{eq:ESLEspinbosonrealtime-1}
\end{equation}
This is essentially the Euler-Maruyama approximation. While the error
of this discretisation is proportional to $\sqrt{\delta_{t/\tau}}$,
it is straightforward to implement directly, unlike more sophisticated
schemes. Provided the timestep is small enough, it has proved sufficiently
accurate for a first implementation of the ESLE. 

\bibliographystyle{unsrt}

\end{document}